\documentclass[12pt,preprint]{aastex}

\renewcommand {\deg}   {\mbox{$^\circ$}}

\newcommand   {\kms}   {\mbox{km\,s$^{-1}$}}
\renewcommand {\ga}    {\mbox{\rlap{\hbox{\lower5pt\hbox{$\sim$}}}\hbox{$>$}}}
\renewcommand {\la}    {\mbox{\rlap{\hbox{\lower5pt\hbox{$\sim$}}}\hbox{$<$}}}

\begin{document}
\pagenumbering{arabic} 
\def\kms {\hbox{km{\hskip0.1em}s$^{-1}$}} 
\voffset=-0.8in

\def\msol{\hbox{$\hbox{M}_\odot$}}
\def\lsol{\hbox{$\hbox{L}_\odot$}}
\def\kms{km s$^{-1}$}
\def\Blos{B$_{\rm los}$}
\def\etal   {{\it et al.}}                     
\def\psec           {$.\negthinspace^{s}$}
\def\pasec          {$.\negthinspace^{\prime\prime}$}
\def\pdeg           {$.\kern-.25em ^{^\circ}$}
\def\degree{\ifmmode{^\circ} \else{$^\circ$}\fi}
\def\ut #1 #2 { \, \textrm{#1}^{#2}} 
\def\u #1 { \, \textrm{#1}}          
\def\nH {n_\mathrm{H}}
\def\ddeg   {\hbox{$.\!\!^\circ$}}              
\def\deg    {$^{\circ}$}                        
\def\le     {$\leq$}                            
\def\sec    {$^{\rm s}$}                        
\def\msol   {\hbox{$M_\odot$}}                  
\def\i      {\hbox{\it I}}                      
\def\v      {\hbox{\it V}}                      
\def\dasec  {\hbox{$.\!\!^{\prime\prime}$}}     
\def\asec   {$^{\prime\prime}$}                 
\def\dasec  {\hbox{$.\!\!^{\prime\prime}$}}     
\def\dsec   {\hbox{$.\!\!^{\rm s}$}}            
\def\min    {$^{\rm m}$}                        
\def\hour   {$^{\rm h}$}                        
\def\amin   {$^{\prime}$}                       
\def\lsol{\, \hbox{$\hbox{L}_\odot$}}
\def\sec    {$^{\rm s}$}                        
\def\etal   {{\it et al.}}                     
\def\la{\lower.4ex\hbox{$\;\buildrel <\over{\scriptstyle\sim}\;$}}
\def\ga{\lower.4ex\hbox{$\;\buildrel >\over{\scriptstyle\sim}\;$}}
\def\refitem{\par\noindent\hangindent\parindent}
\oddsidemargin = 0pt \topmargin = 0pt \hoffset = 0mm \voffset = -17mm
\textwidth = 160mm  \textheight = 244mm
\parindent 0pt
\parskip 5pt

\shorttitle{Sgr A*}
\shortauthors{}

\title{A Three Parsec-Scale Jet-Driven   Outflow  from Sgr A*} 
\author{F. Yusef-Zadeh$^1$, R. Arendt$^2$, H. Bushouse$^3$, W. Cotton$^4$, D. Haggard$^1$, 
M. W. Pound$^5$, D. A. Roberts$^1$, M. Royster$^1$ and  M.  Wardle$^6$}
\affil{$^1$Department of Physics and Astronomy and Center for Interdisciplinary Research in Astronomy,
Northwestern University, Evanston, IL 60208}
\affil{$^2$CREST/UMBC/NASA GSFC, Code 665, Greenbelt, MD 20771} 
\affil{$^3$Space Telescope Science Institute, 3700 San Martin Drive, Baltimore, MD  21218}
\affil{$^4$National Radio Astronomy Observatory,  Charlottesville, VA 22903}
\affil{$^5$University of Maryland, Department of Astronomy,  MD  20742}
\affil{$^6$Dept of Physics and Astronomy,  
Macquarie University, Sydney NSW 2109, Australia}


\begin{abstract} 
The compact radio source Sgr A* is coincident with a 4$\times10^6$ \msol black hole at the dynamical center 
of the Galaxy and is surrounded by dense orbiting ionized and molecular gas. We present high resolution radio continuum images 
of the central $3'$ and report a faint continuous linear structure centered on Sgr A* with a PA$\sim60^0$. The extension of 
this feature appears to be terminated symmetrically by two linearly polarized structures at 8.4 GHz, $\sim75''$ from Sgr A*.  
A number of weak blobs of radio emission with X-ray counterparts are detected along the axis of the linear structure. The 
linear structure is best characterized by a mildly relativistic jet from Sgr A* with an outflow rate 10$^{-6}$ \msol\, 
yr$^{-1}$.  The near and far-sides of the jet are interacting with orbiting ionized and molecular gas over the last 1--3 
hundred years and are responsible for a 2$''$ hole, the ``minicavity", characterized by disturbed kinematics, enhanced 
FeII/III line emission, and diffuse X-ray gas. The estimated kinetic luminosity of the outflow is $\sim1.2\times10^{41}$ erg 
s$^{-1}$, so the interaction with the bar may be responsible for the Galactic center X-ray flash 
inferred to be responsible for 
much of the fluorescent Fe K$\alpha$ line emission from the inner 100pc of the Galaxy.
\end{abstract} 




\section{Introduction and Observations}

Stellar orbits within 1$''$ of the compact radio source Sgr~A* 
at the dynamical center of the Galaxy  are  compelling evidence for a 4 $\times 
10^6$\msol\ black hole (Ghez et al.  2008; Gillessen et al.  2009; Reid and Brunthaler 
2004).  This compact source 
is $\sim$100  times closer to us than the next nearest example of a supermassive black hole 
and presents an unparalleled opportunity to study the processes by which gas is 
captured,  
accreted and ejected from black holes.  
Observations of Sgr A* thus far have   revealed    no hard evidence 
for a  small-scale jet or an accretion disk, and thus the relative ratio 
of the accretion and outflow rates is unknown.  We
have recently reanalyzed archival radio and X-ray continuum and molecular line data that were 
taken with the Very Large Array (VLA) of the National Radio Astronomy Observatory\footnote{The 
National Radio Astronomy Observatory is a facility of the National Science Foundation, operated 
under a cooperative agreement by Associated Universities, Inc.} (NRAO), $Chandra$ X-ray 
Observatory, and the Combined Array for Research in Millimeter-wave Astronomy (CARMA) 
observations  of the inner few pc of Sgr A* to search for 
interaction  sites of a  jet with the surrounding  
material near Sgr A*.


The inner pc of the Galactic center contains  diffuse ionized gas,  
e.g., Sgr A West which orbits   Sgr A* (Lacy et al. 1980; 
Ferriere et al. 2012). Sgr A West 
with its three-arm mini-spiral appearance,  
consists of 
the W arm or arc,  which is 
coupled to the 
Circumnuclear Molecular Ring (CMR), 
and the E and 
N arms of ionized gas. With a  closest projected distance of $\sim 0.1$ pc, Sgr A West shows a bright bar of ionized 
gas, within which a hole or the ``minicavity" with a diameter of 2$''$ is noted.
A chain of circular-shaped radio blob structures links  Sgr A* to the 
``minicavity'' by a ridge of emission at 8.4 GHz (Yusef-Zadeh, Morris \& Ekers 1990; Wardle 
\& Yusef-Zadeh 1992). 
The minicavity shows enhanced FeIII line emission, a high electron temperature
and is kinematically disturbed, 
having  radial velocities ranging between -150 and -290 \kms\,  at its edge 
and a highly blue shifted velocity $\sim -340$ \kms\,  in the interior 
(Eckart et al. 1990; Lacy et  al. 1991; 
Lutz et al. 1993; Roberts et al. 1996; Zhao et al. 2009).  
The kinematic structure may be explained by a shock
due to a hot bubble expanding at  a velocity of 180 \kms\,  (Lutz et al. 1993). 
Here, we present  a faint, highly collimated 
linear structure observed at 
radio wavelengths which is  associated with the ridge of emission linking  Sgr A* to the minicavity. 
We argue that this  is  a synchrotron emitting 
jet  from Sgr A* 
and the blobs are sites of 
dynamical interaction  with  orbiting gas.




The radio data presented here were taken at 4.85 (6 cm), 8.4 (3.6 cm), 15 (2 cm), 
22 (1.3 cm) and 42 (7mm) GHz using multiple configurations of the VLA.  The radio data are 
self-calibrated in phase to remove atmospheric phase errors as well as properly correcting the 
time variability of Sgr A*, by fitting a point source model in the visibility plane. The X-ray 
data derives 
 $\sim1.4\times10^6$ seconds of archival  {\it Chandra} ACIS-I observations
taken between 2000-10-26 and 2011-03-31 which 
were 
recalibrated and combined using CIAO 4.4 (CALDB 4.4.8; Fruscione et al. 2006). 
The merged 0.5--10 keV
image has a resolution of 0.492$''$ and  is normalized at 2.3 keV. 
The SiO data were obtained with 
CARMA 
during the 2009 and 2010 observing
seasons in the D and C array configurations.  The array consisted of 
six 10.4m antennas and nine 6.1m antennas and the maps were made on a
127-point hexagonal mosaic, Nyquist-sampling the 10.4m antenna primary 
beam.  The spatial resolution and
spectral resolutions of the final maps are $8.0\arcsec\times 4.1\arcsec$
and 6.74 \kms, respectively, and the $1~\sigma$ rms noise is 90 mJy/beam.







Figure 1a,b show grayscale 22 and 15 
GHz, high dynamic range ($\sim10^4$),  
images of the inner 30$''$ of Sgr A*. 
We note  a new  linear  feature in both images 
running diagonally crossing the bright N and W arms of 
the mini-spiral, 
along which several blobs (b, c, d, h1 and h2) are detected. 
These   images best reveal  the continuity of the   linear feature along a 
position angle of 
$\sim60^0$ (E of N) extending symmetrically with respect to Sgr A*.  
The linear feature to the SW is thicker and brighter than that of 
the NE  by a factor of 1.5-2, respectively. 
The southern  tip of this continuous 30$''$ 
feature appears to cross the W  arm and becomes 
brighter from 0.2 mJy to about 1 mJy at 22 GHz. 
A bow-shock-like  structure is detected at the position 
where the linear feature crosses the W arm near 
$\alpha, \delta (\rm J2000)\,  17^h\, 45^m\, 38^s.9\, , -29^0\, 00'\, 38''$. 
On the NE  of Sgr A*, blob h1 is extended in the direction of the linear feature. 
Figure 1c shows a
closeup  view of the linear feature at 8.4 GHz within  8$''\times6''$ of  Sgr A* 
where  a ridge of emission  linking  blob $\epsilon$ to Sgr A*. 
The  ridge of emission is also seen as 
a linear feature at  42 GHz in  Figure 1d. We note that the weak linear feature 
at 42 GHz 
extends  for 2$''$ pointing  toward the brightest 
segment of the western rim of the minicavity where 
several compact continuum 
sources are detected.  The morphology of the emission
suggests that the linear feature is impacting the 
mini-cavity. 
A ``dark radio cloud"  labeled on Figure 1d  is interpreted as an embedded
molecular cloud associated with the mini-cavity, as discussed elsewhere, 
showing a deficiency of radio continuum emission.

Unlike total intensity, the polarized emission from the region near Sgr A* is not 
confusion-limited. 
Figure 2a shows the detection of  
linearly  polarized contours of emission superimposed on 
the corresponding total intensity image
at 8.4 GHz.  
The polarized features
SgrA-P1-P4
are  placed to  NE and SW of Sgr A*. 
The brightest  polarized features are  aligned along 
the continuous linear feature, as seen in high resolution images (cf., Fig. 1),  
with the total projected extent of 
$\sim150''\,  (\sim$6 pc).
Figure 2b shows  high resolution polarized 
 image of the 
emission  at 8.4 GHz. 
New polarized sources SgrA-P1 ($17^h\, 45^m\, 43^s.6\, , -28^0\, 
59'\, 22''$), P2 ($17^h\, 45^m\, 43^s.3\, , -29^0\, 00'\, 12''$), 
P3 ($17^h\, 45^m\, 43^s.6\, , -29^0\, 00'\, 08''$) 
located $\sim75'', 45''$ and 60$''$ NE of Sgr A* and 
P4 ($17^h\, 45^m\, 34^s.8\, , -29^0\, 01'\, 09''$)
lying 75$''$ SE of Sgr A* .  
The 
brightest polarized feature P1 shows filamentary structure, extending for $\sim25''$. The lower 
limit to fractional polarization of P1 to P4 at 8.4 GHz   are 15\%, 5\%, 1-2\% and 8\%,  
respectively. 
A common characteristic of polarized features SgrA-P1 to P4 is that they are 
azimuthally extended in the direction perpendicular to the orientation of the linear feature. In 
addition to the P1-P4 polarized sources, we also notice a weak linearly polarized filament with an 
extent of $< 1'$ to the south of Sgr A*, labeled P5 in Figure 2b. 
The extension of the linear feature to the NE and SW 
crosses the concentration of molecular gas associated with the CMR. To examine the possibility 
that polarized features are physically associated with the molecular ring, Figure 2c shows the 
clumpy distribution of SiO (2-1) emission. The linearly polarized features SgrA-P1 to P4 appear 
to be spatially correlated with SiO emission, which is known to be an excellent indicator of 
shocked molecular gas. Polarized features, in particular, appear to be located at the edge of 
molecular clumps. 
P4 appears to be associated with an isolated 
molecular clump traced by HCN line emission (Christopher et al. 2005). 
The spectrum of the HCN  clump at the position of P4 
shows two velocity components at --43 and --120 \kms. 


Although there is no polarized emission detected from Sgr A* at a level of less than 0.1\%, we notice weak 
polarized emission at a level of 50-100 $\mu$Jy within the inner few arcseconds of Sgr A*. Figure 2d shows 
contours of 8.4 GHz emission superimposed on diffuse polarized emission with a fractional polarization 
$\sim0.5$\% outside Sgr A*. The diffuse polarized emission is elongated along the axis that links P1 and P4 
and the overall electric vector distribution of weak polarized emission outside Sgr A* runs perpendicular to 
the axis of the elongation. 
Using  the upper limit flux density of $\sim1$ mJy per 0.35$''\times0.2''$ beam,   the equipartition 
magnetic fields are 1.6, 1.9 and 5.9 mG for proton to electron ratio of 0, 1 and 100, respectively. 




Figure 3a shows a 8.4 GHz image of the inner $\sim45''\times30''$ of 
Sgr~A*. The bright central source 
coincides with Sgr A* surrounded 
by 
the three  arms of the mini-spiral and  the bar of ionized 
gas within  which the ``minicavity'' lies. 
 SgrA-a1, a2, b, c, d and e,  are identified to the west of Sgr A* crossing the ionized 
flow associated with the W  arm. 
The $\epsilon$ blob,   which was originally detected at 15 GHz (Yusef-Zadeh, Morris and Ekers 1990), 
is the brightest of all the blobs and lies about 1$''$ to the west of 
Sgr A*, followed by two blobs a1 and a2 lying at the western boundary of the minicavity. 
The SgrA-blobs f, g and h are detected to the east of Sgr A* 
crossing the N arm of Sgr A West. 
The peak flux density of 
the blobs a1 and a2 range between 0.1 to 0.3 mJy 
beam$^{-1}$. 
Figure 3b shows a 22 GHz image 
displaying  diffuse emission lying between the W  arm 
and  the minicavity.  A weak circular-shaped feature 
 surrounding 
blob a1 and a2  with an X-ray counterpart 
(c.f. G359.942-0.045 in Fig. 5) is detected with a diameter of 0.33$''$. 
The integrated 
flux density of the circular-shaped 
feature which lies immediately adjacent to the minicavity,  is $\sim$29 mJy 
(cf, Figs.  4a and 5a). 



Figure 4a,b show a 22 GHz VLA and 2-4 keV Chandra image of the inner 25$''\times18''$ of Sgr A*, 
respectively. One of the new radio features that is revealed in Figure 4a is a $\sim2''$ tear-drop shaped 
bubble roughly centered around Sgr A*.  This bubble is open on the northern side
with a morphology that 
resembles the shape of the minicavity. Detailed analysis of Chandra data suggests that X-ray emission from 
Sgr A* is spatially extended (Baganoff et al. 2003). However, the coincidence of the extended X-ray 
emission from Sgr A* with the tear-drop suggests that the extended X-ray component 
is likely produced by the synchrotron jet from Sgr A* or 
shocked winds   by mass-losing stars  orbiting Sgr A*, as the ram pressure 
of the winds pushing the ionized bar away and  create   a tear-drop cavity .

There are  four  diffuse and compact  X-ray sources 
that have previously been 
identified,  
 Sgr A*, IRS 13, G359.945-0.044 (a PWN  candidate) and 
G359.943-0.047 (Baganoff et al. 2003; Wang et al. 2006; Muno et al. 2008, 2009), as labeled in 
Figure 4a,b. 
There are two faint extended X-ray sources symmetrically placed with 
respect to Sgr A*. 
One is an elongated loop-like structure (labeled NE plume in Figure 4) to 
the NE of Sgr A* with an extension of 8$''\times2''$. The SW plume 
is more difficult to discern because of  
the bright source  IRS 13 and the transient source G359.943-0.047.  
However, 
the X-ray source G359.942-0.045, which has a radio counterpart (blob b in Figures 1 and 4),  
appears  to be a part of the SW  plume. 
The two extended plume-like features lie along the axis in 
which the continuous linear structure is detected at radio wavelengths. 
There are also weak X-ray features noted in the immediate vicinity of 
the N arm, G359.947-0.047 in Figure 4b,  
along the same axis as the linear radio feature. The X-ray emission 
adjacent to the N arm provides a strong evidence that an energetic event is 
responsible for X-ray emission. 


Our  high resolution radio and X-ray continuum images show new structural details 
within 5 pc of Sgr A*. We detect a chain of faint blobs which are 
distributed along a continuous linear feature. The linear feature continues beyond the ionized 
arms of Sgr A West, extending to about 75$''$ (~3pc) away from Sgr A* before it is terminated by 
extended linearly polarized features.
The orientation  of the collimated feature is consistent with proper motion 
and cometary morphology 
of dust features as well as the bow shock structure of FeII line emission 
leading  to the idea of an outflow from young massive stars or 
from Sgr A* (Lutz et al. 1993; Muzic et al. 2007). 
 The collimated feature 
could be produced by  the winds of massive young 
stars associated with the clockwise rotating stellar disk (Paumard et al. 2006). 
However, it is not clear how to 
collimate the outflowing winds.  Numerical simulations of wind-wind collisions incorporating  
the dynamics of stars in the central cluster do not show any evidence of collimated
 structure (Cuadra et al. 2008). 
Thus, it is 
more likely that Sgr A* itself is responsible for the origin of the collimated feature and the 
acceleration of particles to high energies.

We account for the origin of the 
new features in terms of a mildly relativistic jet symmetrically emanating from Sgr A*. The jet 
interacts with the orbiting  gas, forms the structure of the minicavity, compresses and shocks the gas
producing X-ray and radio emitting  blobs of  emission. 
The 
brightest polarized features are symmetrically placed at a projected distance of $\sim$3pc from 
Sgr A* and are consistent with the jet picture. 
The lack of polarized emission 
at 8.4 GHz from the inner pc 
is due to depolarization by dense ionized gas.

The near and far side components of the symmetrical jet run into the ionized bar and the N arm 
which are assumed to be located in the front and back side of Sgr A*, respectively. 
In this 
picture, the ionized bar and the N arm are two independent ionized features orbiting Sgr A* 
(Zhao et al. 2009). 
The ram pressure of the near-side component  of the jet 
disturbs the  ionized bar and produces the 
high negative velocity at the western edge of the minicavity. 
The change in the velocity of the gas is about 100 \kms\,  between the E and W 
boundaries of the minicavity. The blue-shifted gas associated with the bar gets more 
blue-shifted at the interaction site. Similarly, the far side component of the jet punches 
through the N arm and causes a sinusoidal pattern along the clockwise motion of the N arm.  If 
we assume that the jet is mildly relativistic, it 
can punch through the 10$^5$ cm$^{-3}$ gas of the bar.  On the other hand if the jet missed the bar originally and then the bar 
is currently sweeping around it, the jet can survive deflection and the interaction might create 
the minicavity.  


The mass outflow rate in the jet should be somewhat less than the accretion rate onto Sgr A*, 
which is estimated to be a few times 10$^{-5}$ \msol\, yr$^{-1}$ (e.g. Cuadra et al. 2008).  
We adopt a nominal mass outflow rate of 10$^{-6}$ \msol\, yr$^{-1}$, a Lorentz factor 
$\gamma\sim$3, and a cross-section equivalent to a diameter of 1".  Then the momentum transfer 
rate $\dot{M}\gamma$v = 0.3 $\gamma$ \msol\, yr$^{-1}$ km s$^{-1}$, sufficient to drive a 
270$\gamma^{0.5}$ \kms\, shock into the 10$^5\, \rm cm^{-3}$ molecular gas associated with the 
bar.  The interaction of this gas with the jet may then be responsible for the minicavity, 
which exhibits a velocity disturbance of 100-200 \kms\, (Roberts et al. 1996), suggesting an 
age of approximately 100-300 years for this interaction. 
The gas would be 
shock-heated to about 3 million K, with a cooling time of only a few years, and the X-ray 
luminosity of the cooling gas is of order the kinetic luminosity of the jet.  
If the interaction only lasted for 1-2 decades, it would have produced the X-ray flash thought 
to be responsible for much of the fluorescent 6.4 keV Fe K$\alpha$ line
emission within 100 pc of the Sgr A* (Sunyaev \& Churazov 1998; Koyama et al. 2009).

The jet continues crossing the W and N arms and eventually  interacts  with the molecular ring 
where linearly polarized emission is detected. The wider thickness and higher surface brightness 
 of the jet to the SW, 
compared to that of NE are  consistent with a higher density of material distributed in 
the dense ionized bar. The jet drives  shocks into gas clouds producing blobs of X-ray emitting 
gas, shocked molecular H$_2$ and SiO emission, enhancing the iron abundance in the gas phase, 
generating a low dust to gas ratio in the minicavity and disturbing the kinematics of orbiting 
ionized gas.

The collimation of the outflowing material from Sgr A* along a jet axis is generally considered 
to be due to an accretion disk.  The axis of the jet outflow from Sgr A* 
is estimated to be 
60$^0$.  This position angle of large-scale jet differs by 30$^0 -60^0$ 
from  the constraints placed by VLBA measurements (Markoff, Bower and Falcke 2007). 
Notably, the orientation  of the jet, indicated  by dashed 
line in schematic diagrams of Figure 4c,d
 runs perpendicular to the principal axis of  the clockwise stellar disk (Paumard 
et al. 2006).  
This alignment may be a coincidence or possibly due to the accretion processes that 
still continue in forming young low-mass stars. 
 If so, Sgr A* is  retaining  the memory of  in-situ 
star formation activity in its vicinity. 
Almost all models of the steady component of the emission from Sgr A* predict an accretion disk, 
a jet and/or an outflow. 
Present observations support jet models for   Sgr A* 
(e.g., Falcke and Markoff 2000; Das et al. 2009; Becker et al. 2011). 



In summary, we have presented evidence for a faint pc scale linear feature  arising from 
Sgr A* that  interacts with ionized and molecular material 
orbiting Sgr A*. 
An important implication of a jet/outflow is that a disk is needed to  
collimate   the outflowing materials.  
Given that the VLA is a linear array, deconvolution    
errors may result linear structures along the symmetric 
beam  pattern with the beam  being 
strongest in the NS direction for southern sources. 
Our images do not show 
any evidence of symmetric  ``cross''  pattern. 
However, future observations with non-linear arrays,  such as ALMA,  should 
confirm  the reality of polarized jet associated with Sgr A*, as 
reported here.

Acknowledgments:
This work is partially supported by the grant AST-0807400 from the NSF and 
DP0986386 from the Australian Research Council.
Ongoing CARMA development
and operations are supported by the National Science Foundation under a
cooperative agreement, and by the CARMA partner universities.





\begin{figure}[p]
\includegraphics[scale=0.35,angle=0]{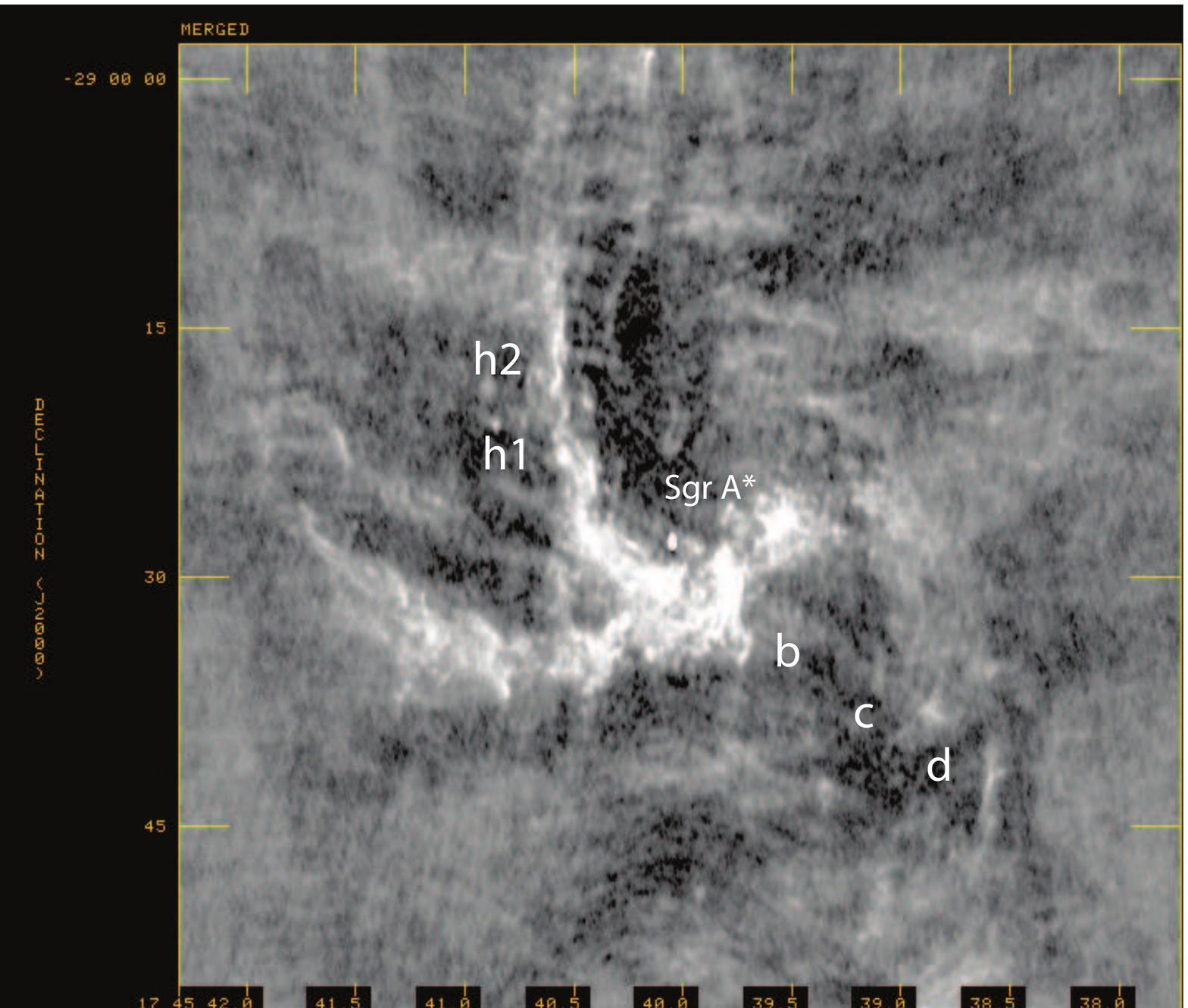}
\includegraphics[scale=0.35,angle=0]{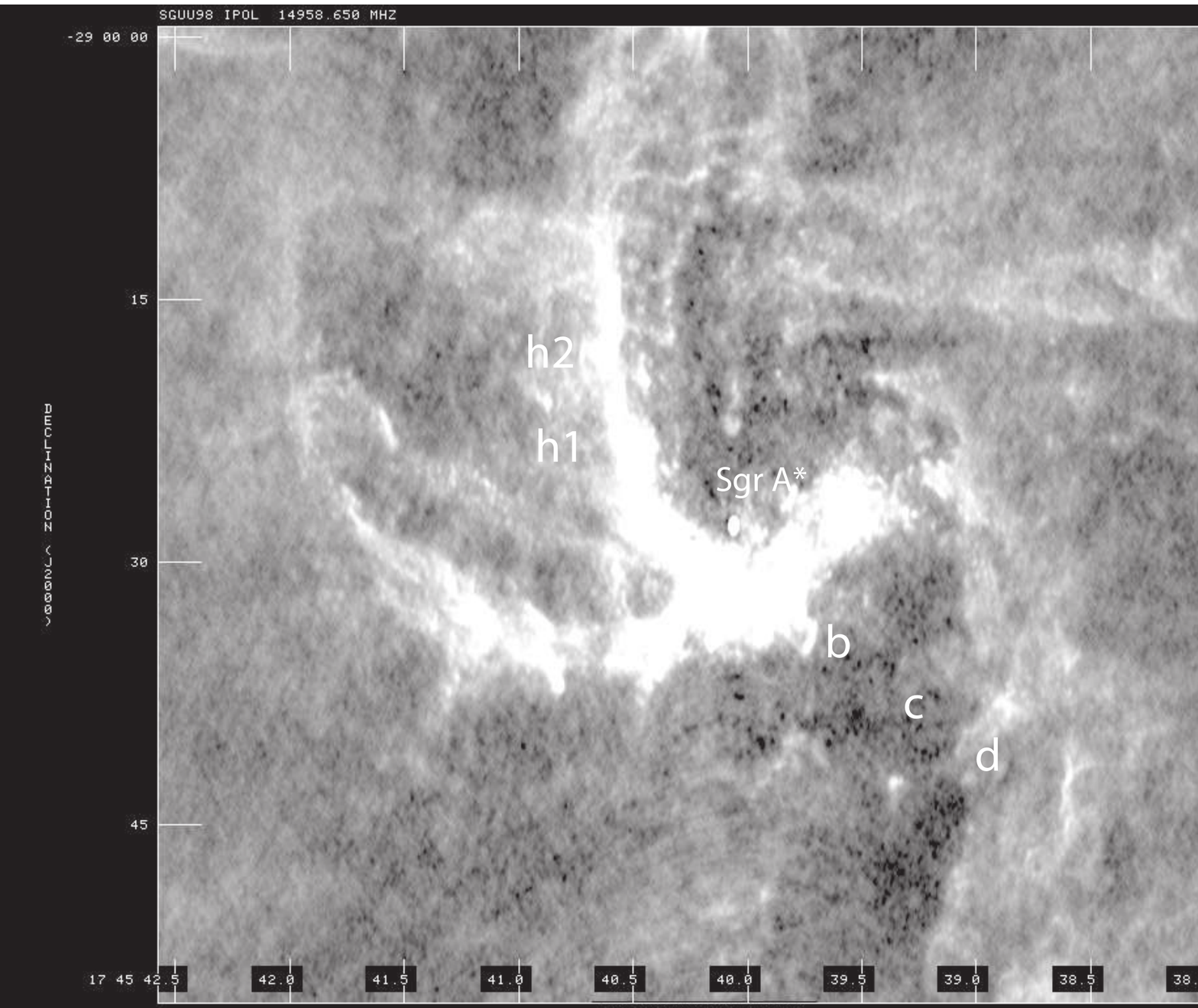}
\includegraphics[scale=0.35,angle=0]{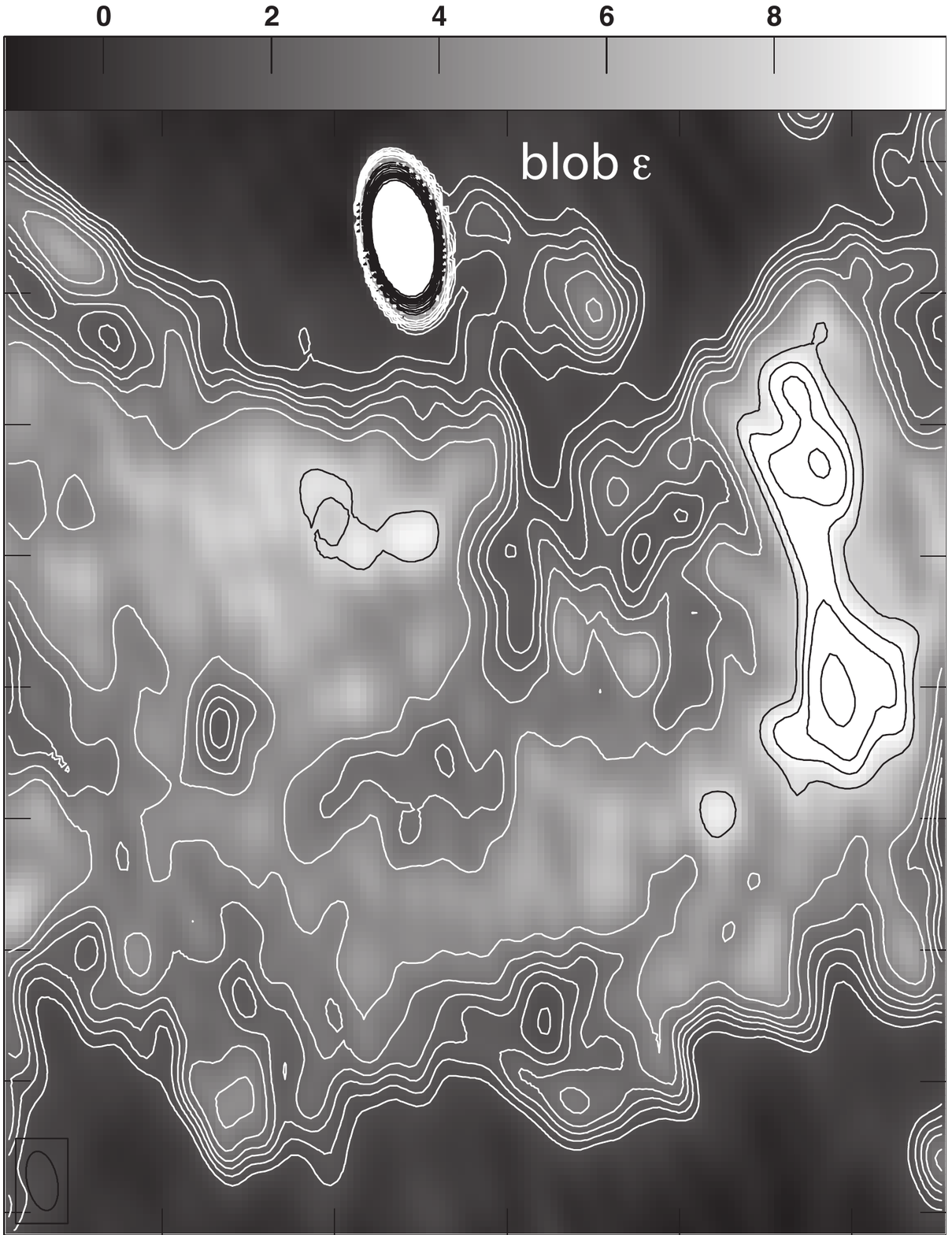}
\includegraphics[scale=0.35,angle=0]{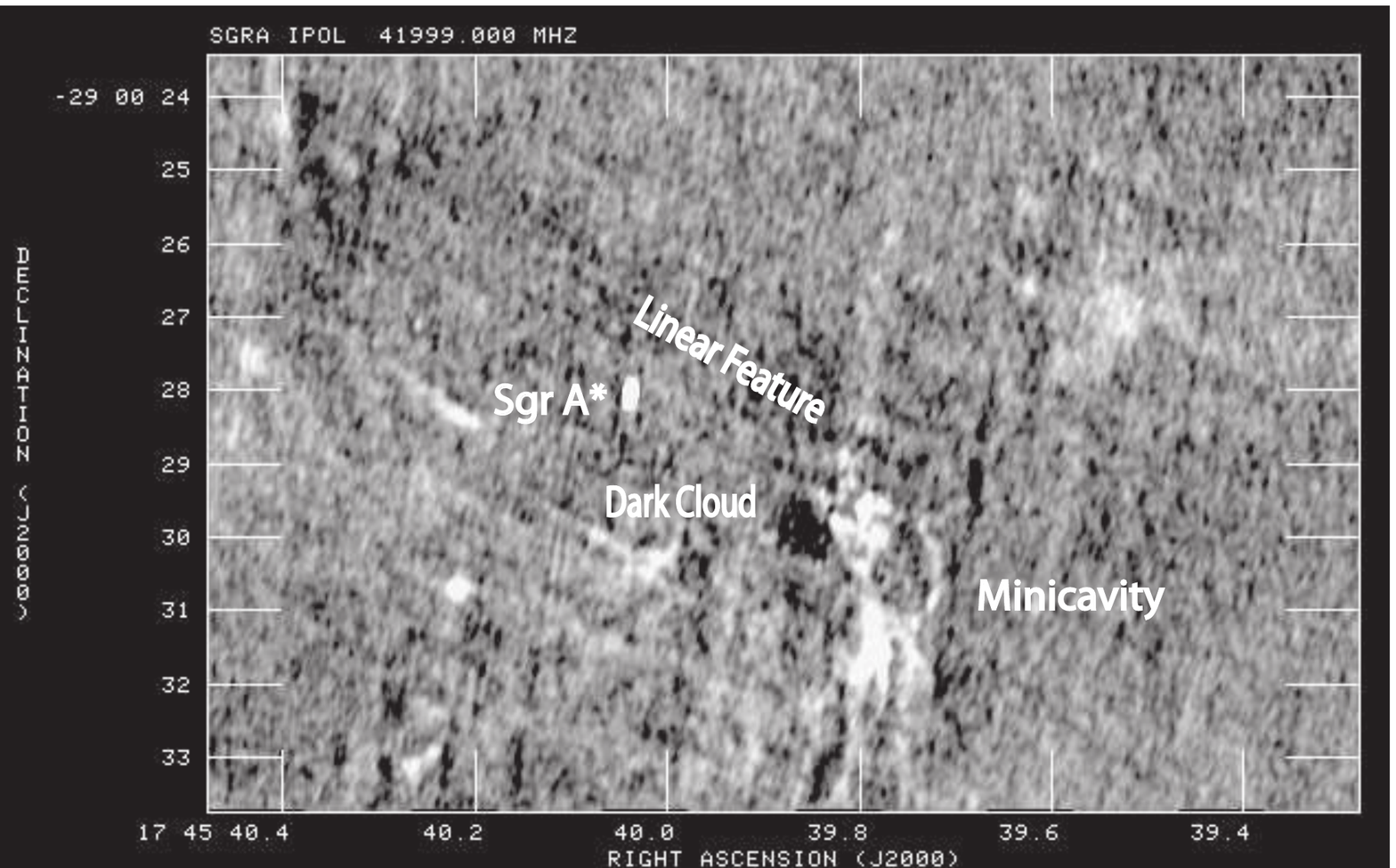}
\caption{
{\it (a-b) Top}
22 (left) and 15 (right) GHz VLA images  constructed by 
combining A and B array data 
with resolutions  of 
$0.36''\times0.18''$ (PA=2$^0$) and 
$0.36''\times0.20''$ (PA=-0.6$^0$), and rms noises of 1.5$\times10^{-4}$ and 
9$\times10^{-5}$ mJy beam$^{-1}$,  respectively.
{\it (c) Bottom Left} 
Contours of emission with a resolution of $0.46''\times0.23''$ (PA=-12$^0$) 
at 8.4 GHz  -1.2,  1.2, 1.6, 2, 2.4, 3,  4,  6, 8, 10, 12, 16, 20,...,  40 and  50 mJy beam$^{-1}$.
{\it (d) Bottom Right} 
A  42 GHz image with  a resolution $0.16''\times0.08''$ (PA=-0$^0$). 
}
\end{figure}

\begin{figure}[p]
\includegraphics[scale=0.4,angle=0]{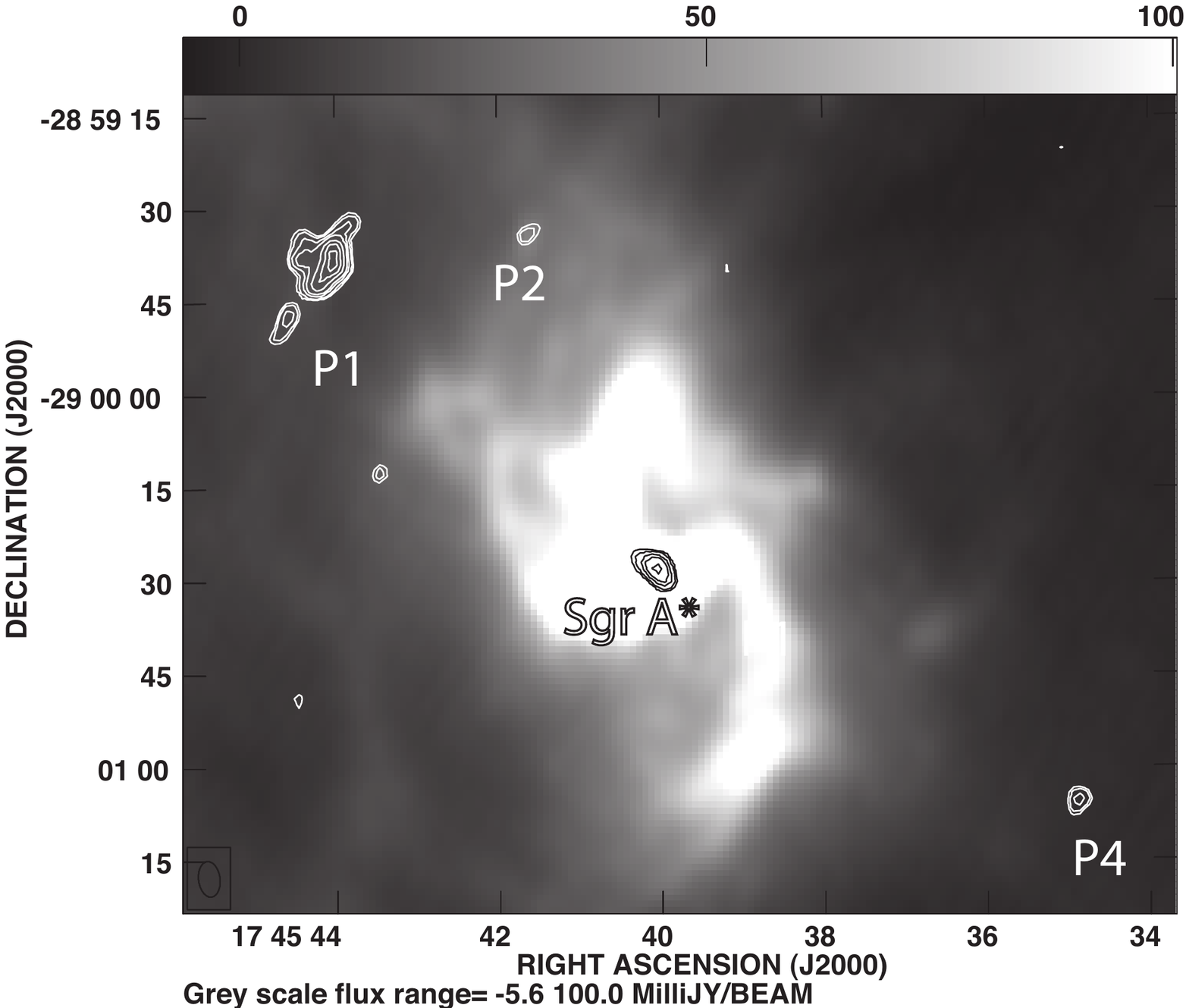}
\includegraphics[scale=0.3,angle=0]{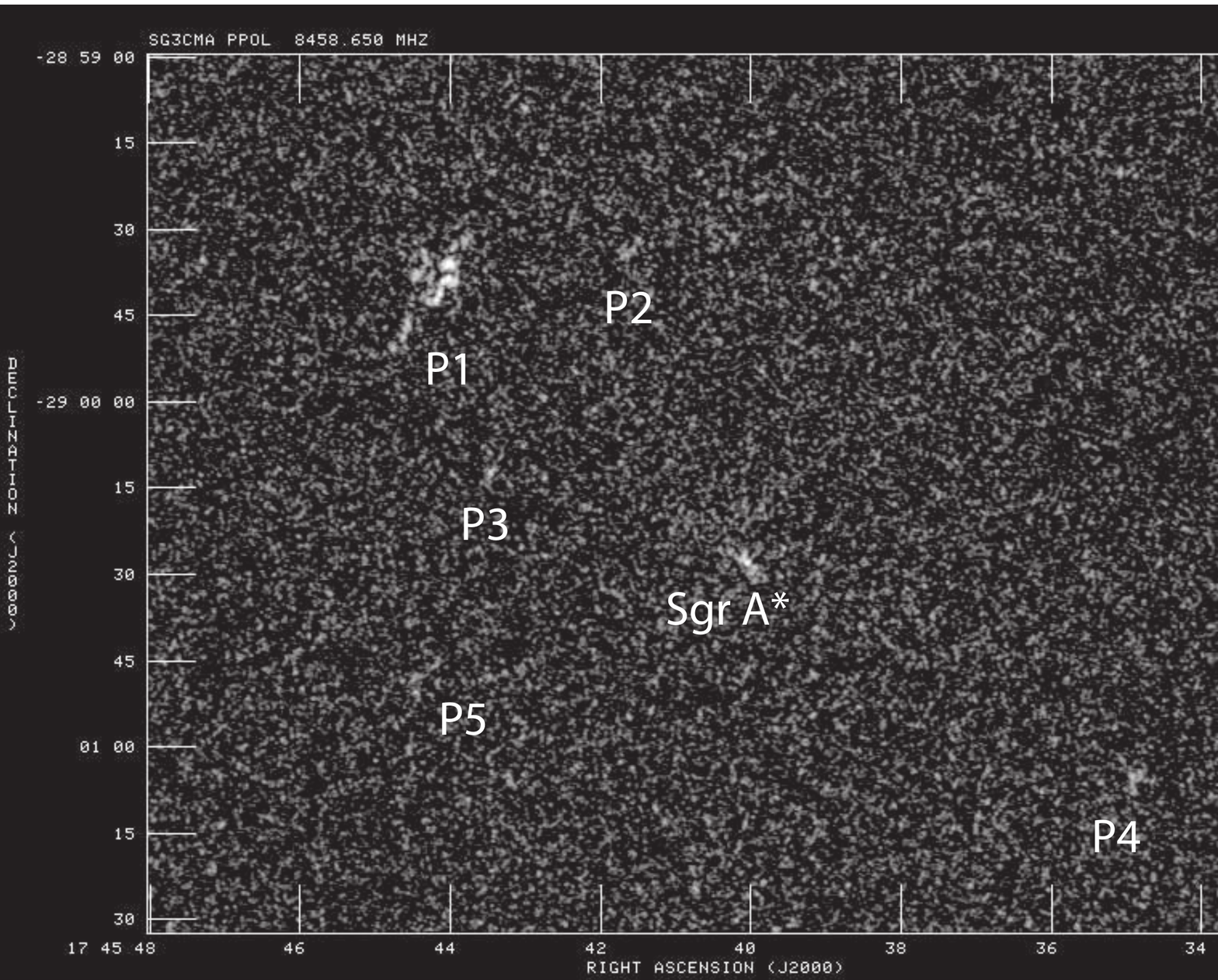}\\
\includegraphics[scale=0.3,angle=0]{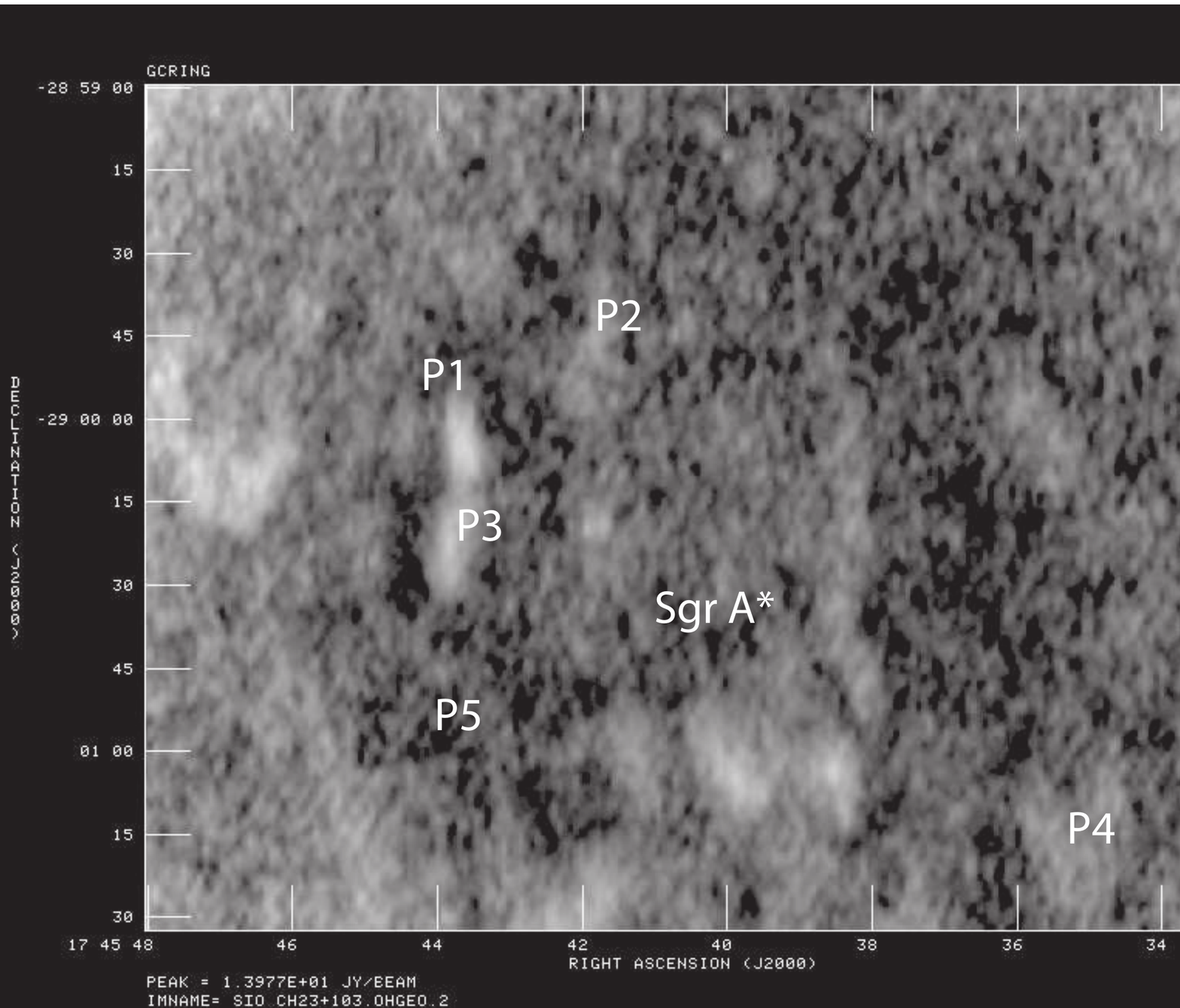}
\includegraphics[scale=0.35,angle=0]{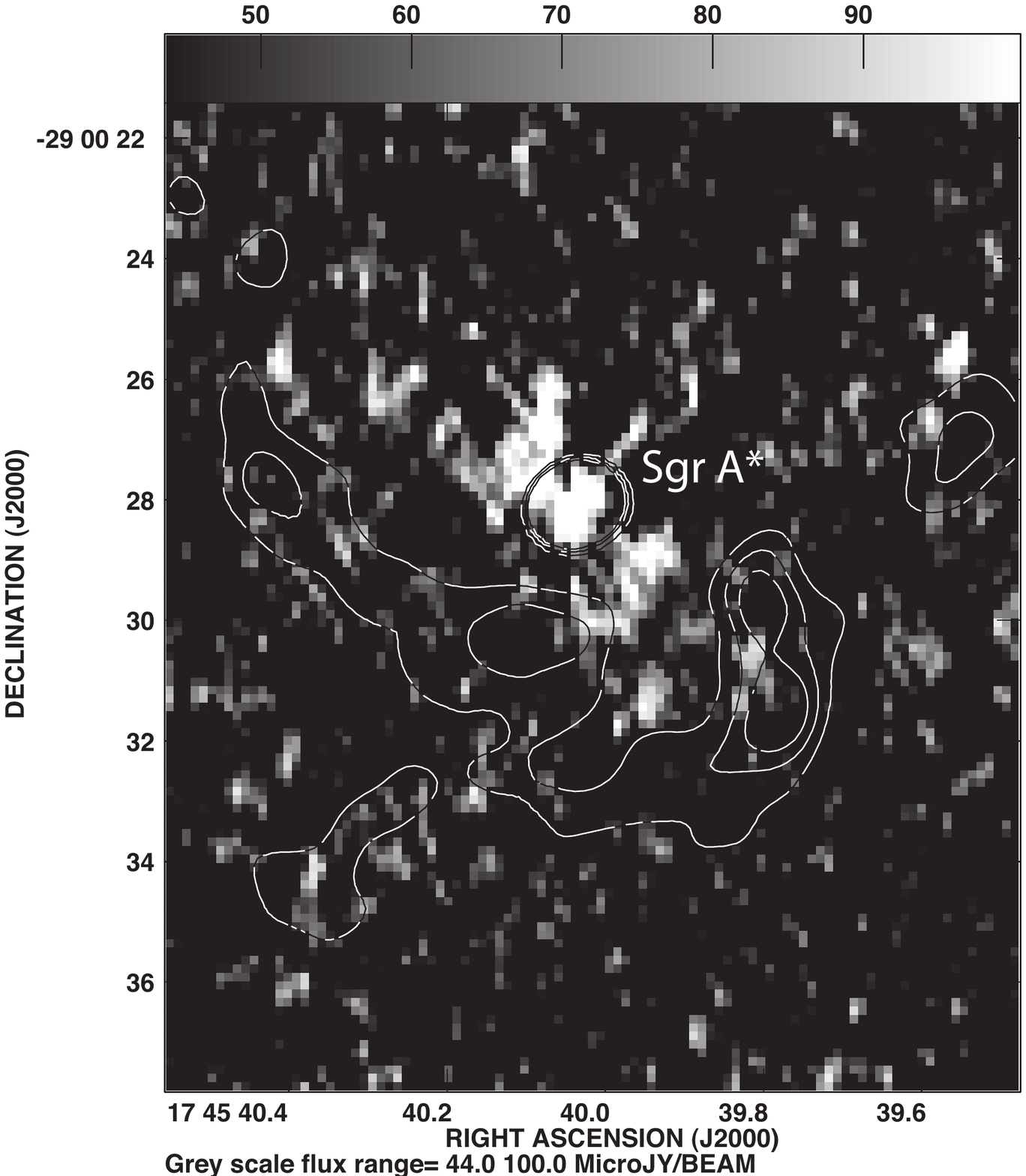}
\caption{
{\it (a)} 
Contours of 8.4 GHz polarized emission 
with a resolution  of 
10$''\times10''$ with levels set at 
(3.5, 4, 5, 6, 8, 10, 12)$\times50\mu$Jy beam$^{-1}$ 
on a grayscale intensity image with a resolution of 
$5.8''\times3.5'' (PA=7^0$). 
{\it (b)} 
A grayscale 8.4 GHz  polarized  
intensity at 8.4 GHz 
with a resolution of 
0.93$''\times0.51''$\,  (PA=-7$^0$). 
{\it (c)} 
A grayscale distribution of SiO(2-1) emission from the CMR integrated between 
velocities -121.6 and 148 \kms. 
{\it (d)} 
Contours of 8.4 GHz intensity of the minicavity 
with levels set at 20, 30 and 40 mJy  beam$^{-1}$
superimposed on polarized intensity from Sgr A* and its immediate vicinity. 
The resolution is the same as (b). 
} 
\end{figure}

\begin{figure}[p]
\includegraphics[scale=0.35,angle=0]{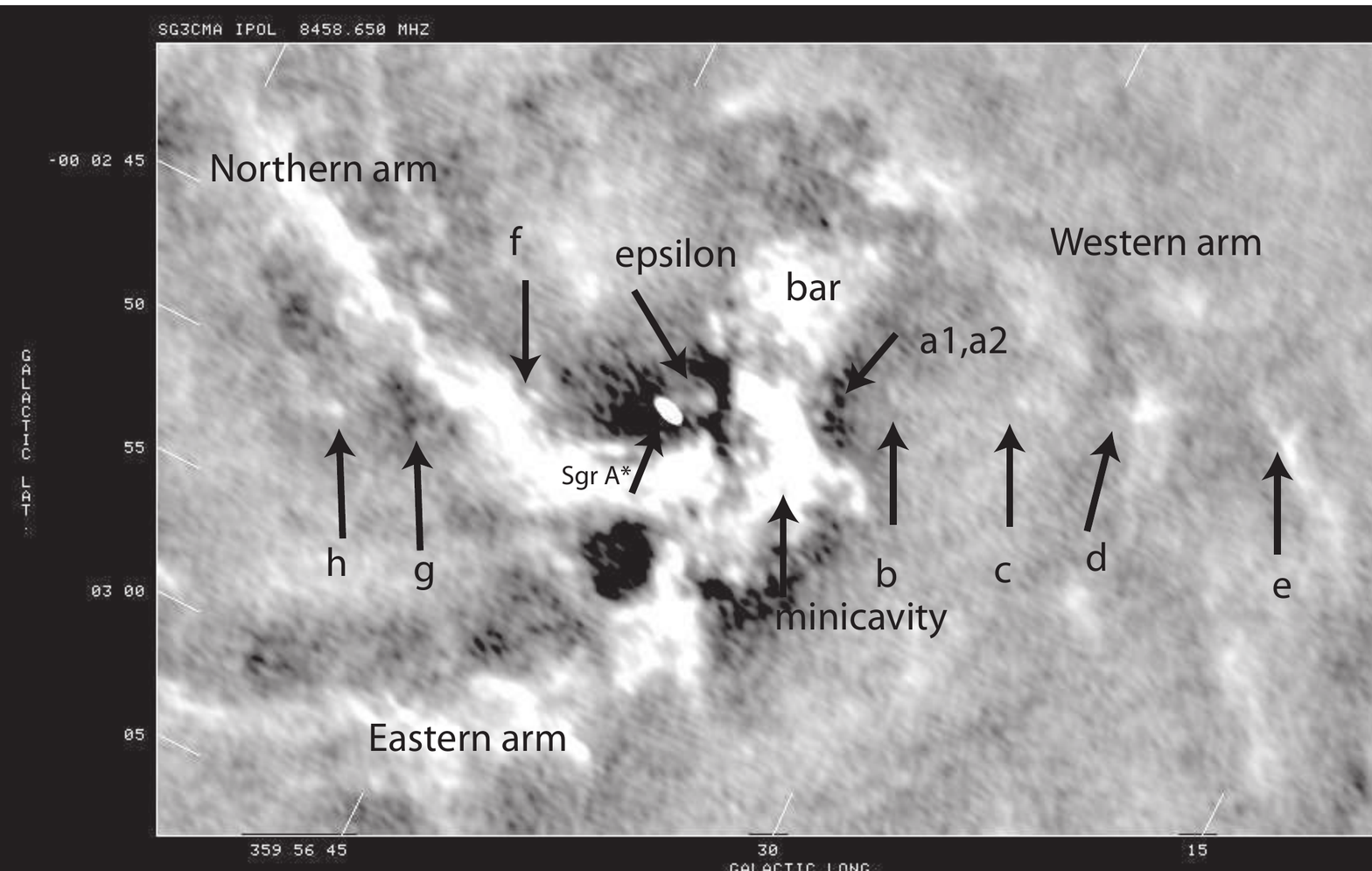}
\includegraphics[scale=0.35,angle=0]{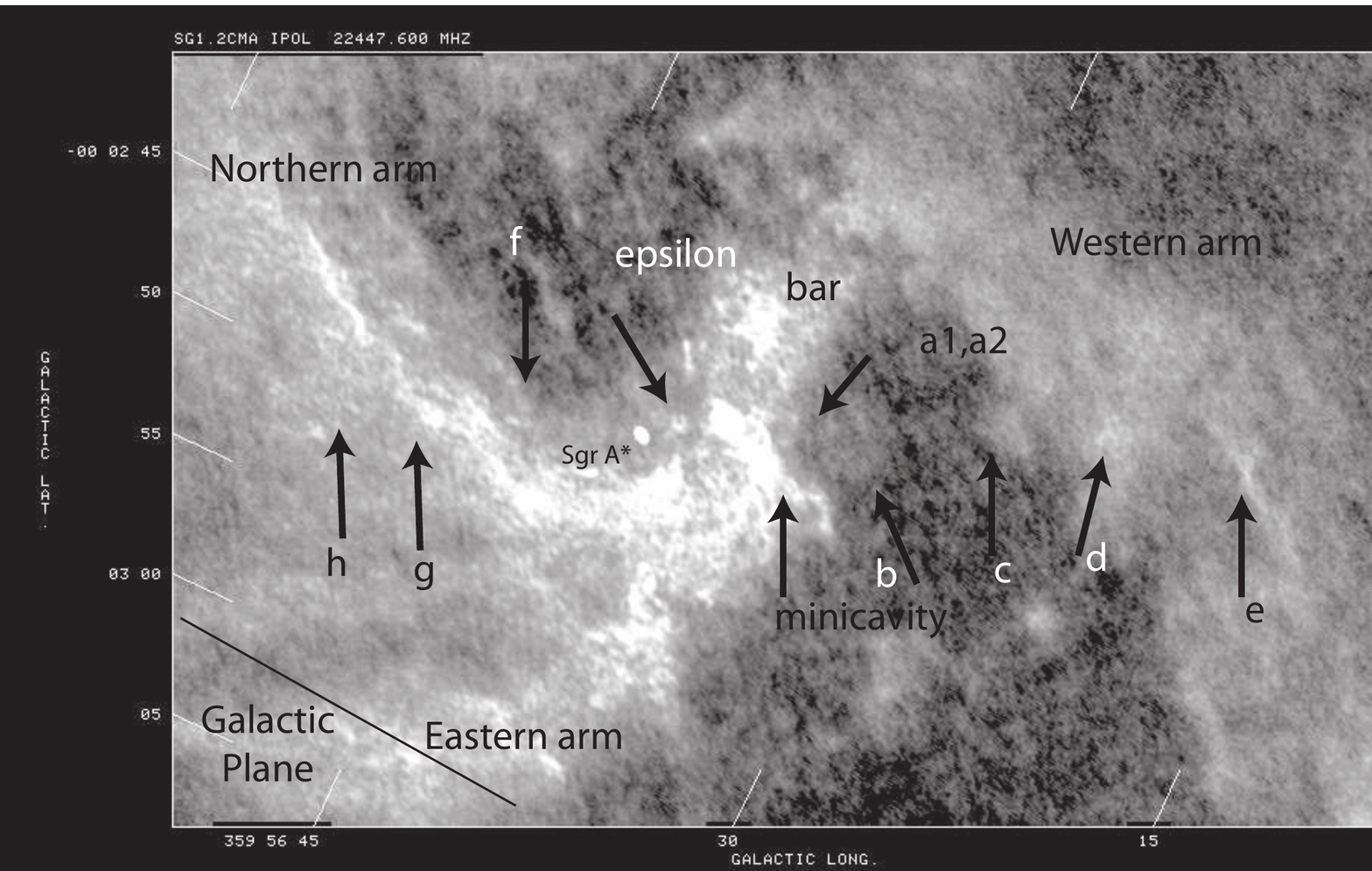}\\
\includegraphics[scale=0.4,angle=0]{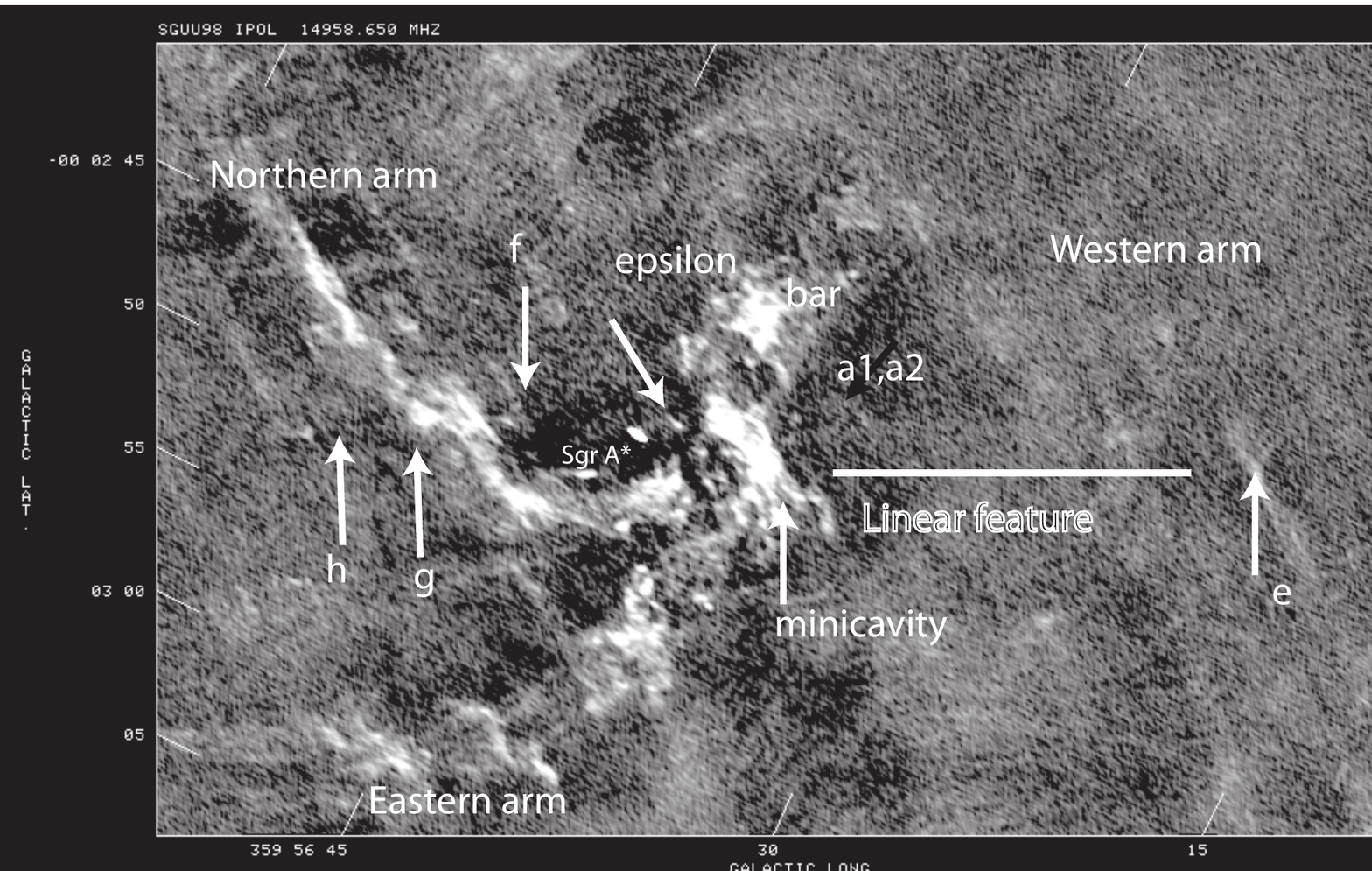}
\caption{
All  images are rotated by $\sim28^0$ with respect to the Galactic plane
in order to place the linear feature along the horizontal axis. 
{\it (a)} 
A grayscale total intensity image of the inner 
30$''\times45''$ of Sgr A* at 8.4 GHz based on 
combining A-array configuration data sets  taken on 1991-08-15
and 1990-05-30. The spatial resolution 
is  $0.46''\times0.23''$  (PA=$12^0$).
{\it (b)} 
Similar  to (a) except  at 23 GHz based on combining 
A, C and B-array data taken on 1990-05-30, 1986-12-28
and 1987-11-27, respectively. 
The resolution of the image is  0.24$''\times0.17''$ (PA=$1.58^0$). 
{\it (c)}
Similar  to (a) except at 15 GHz based on 
A-array  data taken
on 1998-04-30. The resolution is 
0.2$''\times0.1''$ (PA=$-0.23^0$). 
} 
\end{figure}


\begin{figure}[p]
\centering
\includegraphics[scale=0.75,angle=0]{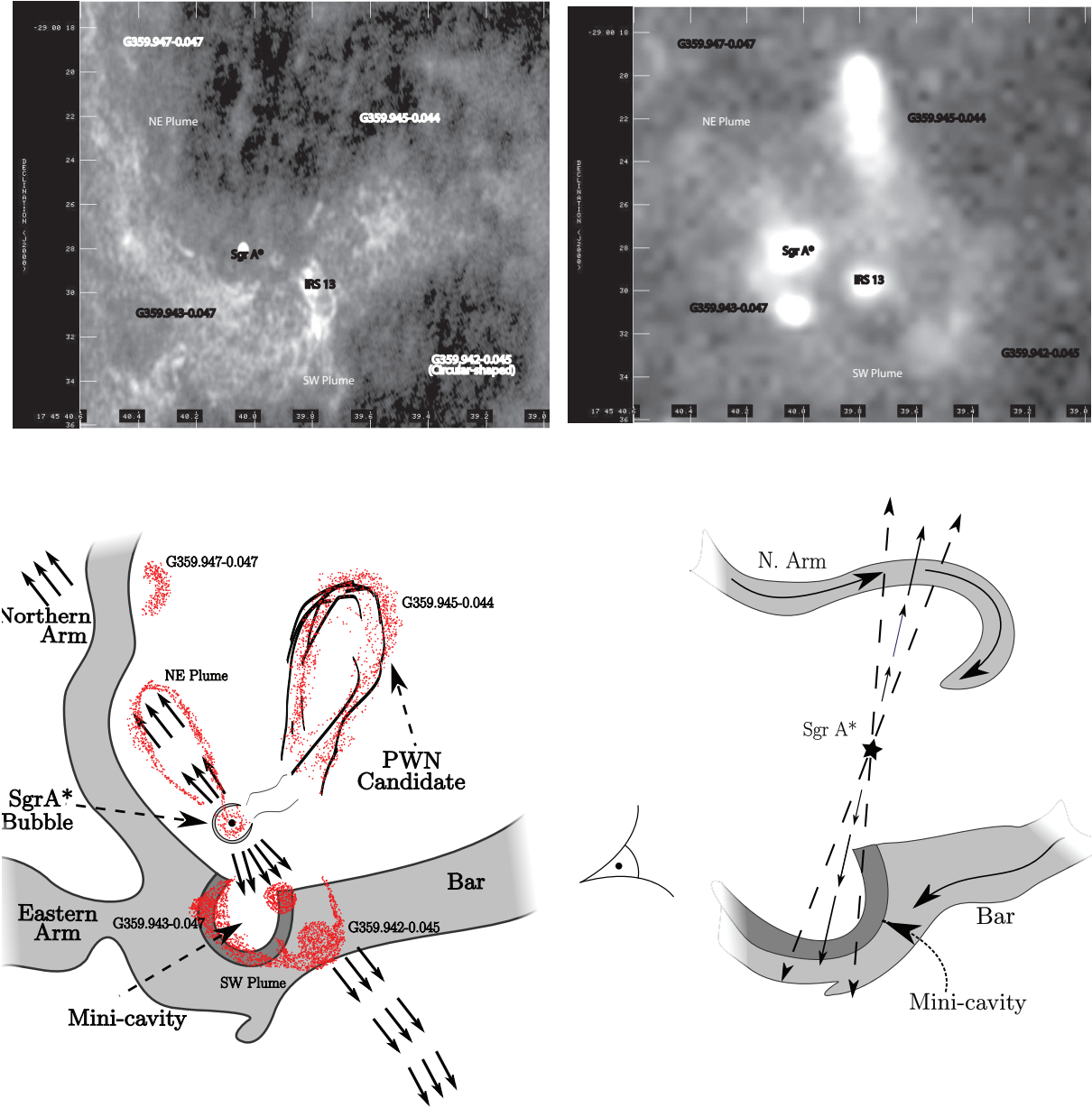}
\caption{
{\it (a)} 
A radio continuum image of Sgr A* at 22.3 GHz with a spatial resolution of 
$\sim0.24''\times0.17''$ (PA= 1.6$^0$). 
The data are taken in the A-array configuration on 1990-05-30, 1987-11-27, 
combined  with the C-array configuration on 
1986-12-28.   
{\it (b)} 
The same region as  (a) but 
showing X-ray emission 
between 0.5-10keV at  a resolution of 0.49$''$. 
{\it (c)} 
A schematic diagram showing bright 
radio and X-ray features in  gray  and red, respectively, and 
a model of a narrow two-sided jet from Sgr A*. 
(d)  A schematic diagram of a model 
showing 
the near and far side components of the 
symmetrical jet running  into the ionized bar and the N arm 
which are assumed to be located in the front and back side of Sgr A*, respectively. 
}
\end{figure}
\end{document}